\documentclass[11pt]{article}
\usepackage{amsmath,amssymb,color,graphics,epsfig,cite}
\usepackage{amsfonts}
\newcommand{\hoch}[1]{$\, ^{#1}$}

\newcommand{\be}{\begin{equation}}
\newcommand{\ee}{\end{equation}}
\newcommand{\bea}{\setlength\arraycolsep{2pt} \begin{eqnarray}}
\newcommand{\eea}{\end{eqnarray}}
\newcommand{\nn}{\nonumber}

\def\ft#1#2{{\textstyle{\frac{\scriptstyle #1}{\scriptstyle #2} } }}
\def\fft#1#2{{\frac{#1}{#2}}}

\def\0{{\sst{(0)}}}
\def\1{{\sst{(1)}}}
\def\2{{\sst{(2)}}}
\def\3{{\sst{(3)}}}
\def\4{{\sst{(4)}}}
\def\5{{\sst{(5)}}}
\def\6{{\sst{(6)}}}
\def\7{{\sst{(7)}}}
\def\8{{\sst{(8)}}}
\def\sst#1{{\scriptscriptstyle #1}}

\begin{document}

\begin{center}
{\Large {\bf Vulnerability of Horizon Regularity: Horizon as a Natural Boundary}}

\vspace{20pt}

Qi-Yuan Mao\hoch{1}, Liang Ma\hoch{1} and H. L\"u\hoch{1,2}

\vspace{10pt}

{\it \hoch{1}Center for Joint Quantum Studies and Department of Physics,\\
School of Science, Tianjin University, Tianjin 300350, China }

\bigskip

{\it \hoch{2}Joint School of National University of Singapore and Tianjin University,\\
International Campus of Tianjin University, Binhai New City, Fuzhou 350207, China}
\vspace{40pt}

\underline{ABSTRACT}
\end{center}

We consider Einstein gravity extended with Riemann-squared term and construct the leading-order perturbative solution to the rotating black hole with all equal angular momenta in $D=7$. We find that in the extremal limit, the linear perturbation involves irrational powers in the near-horizon expansion. We argue that, despite that all curvature tensor invariants are regular on the horizon, the irrational power implies that the inside of the horizon is destroyed and the horizon becomes the natural boundary of the spacetime. We demonstrate that this vulnerability of the horizon regularity is an innate part of Einstein theory, and can arise in Einstein theory with minimally coupled matter. However, in fine-tuned theories such as supergravities, the black hole inside is preserved, which may be one of the criteria for a consistent theory of quantum gravity. We also show that the vulnerability occurs in general higher dimensions, which only a few sporadically distributed dimensions can evade.

\vfill{qiyuan\_mao@tju.edu.cn\ \ \ liangma@tju.edu.cn\ \ \  mrhonglu@gmail.com}

%{\footnotesize \hoch{*}Corresponding author}

\thispagestyle{empty}
\pagebreak

\tableofcontents
\addtocontents{toc}{\protect\setcounter{tocdepth}{2}}

\newpage

\section{Introduction}

Black holes, predicted by Einstein's theory of General Relativity, are characterized by having an event horizon, a boundary where the escape velocity equals the speed of light. In all the well-known black holes such as the Schwarzschild or the Kerr black holes, in four or higher dimensions, the curvature invariants on the horizon are regular and the near-horizon geometry is described by analytic metric functions that are infinitely differentiable. (see {\it e.g.}~\cite{Kerr:1963ud,Carter:1968ks,Myers:1986un,Hawking:1998kw,Gibbons:2004js,
Gibbons:2004uw,Chen:2006xh,Lu:2008js} for some Ricci-flat or Einstein metrics.) This implies that the horizon is not the boundary of spacetime itself. One can extend the horizon geometry, {\it e.g.}~\cite{Kruskal:1959vx} to include the inside of a black hole and indeed this is confirmed by analysing the geodesic motion.

Recently, the linear perturbation of higher-derivative gravity to $D=5$ dimensional rotating black holes with equal angular momenta was obtained \cite{Ma:2020xwi}. In the extremal limit, the perturbed solution involves a function like $(c+\log(1-r_0/r))$, which is no not analytic on the horizon $r=r_0$. In this particular case, the ${\rm i} \pi$ factor generated by crossing over the horizon can be absorbed into the integration constant $c$. This, however, leads to some obvious questions: can worse non-analyticities arise on the horizon and what are their implications?

Black hole extremal limit typically arises when the attractive force of gravity balances the repulsive force in the theory. The best known example of such balance is exhibited by the Reissner-Nordstr\"om (RN) black hole, where gravity and electric Coulomb repulsion balance precisely in the extremal limit. This can not only be verified by geodesic motion of charged test particles \cite{Zhao:2018wkl}, but also allows the construction of multi-center black holes in harmonic superposition. This no-force condition was recently shown to be preserved under appropriate higher-order curvature corrections \cite{Hu:2023qhs}. Rotations provide a repulsive centrifugal force and extremal rotating black holes are the results of its balance against the gravitational attraction. However, in higher dimensions, gravitational force becomes weaker, whilst the centrifugal force, associated with rotation on a plane, remains the same strength. This implies that the balance between gravity and centrifugal force becomes more strenuous in higher dimensions so that any perturbation can ruin the near-horizon geometry.

In this paper, we therefore consider Einstein gravity extended with a Riemann-squared term in general odd $D=2n+1$ dimensions, with a coupling constant $\alpha$. We construct the leading $\alpha$-order perturbation to the Ricci-flat rotating black holes. For simplicity,  we consider only the cohomogeneity-one metrics with all equal angular momenta. We find that in the extremal limit, the linearized solutions in general higher odd dimensions involve a non-analytic term of the type $(r-r_0)^{\Delta_+}$, where $\Delta_+$ is an irrational number. In fact, we find that it is generally irrational in higher $D=2n+1$ dimensions lying between 1/2 and 1, but it can be rational sporadically.

    We study the implication of these irrational structures and we argue that despite that
curvature tensor invariants are all regular and geodesics are incomplete on the horizon, we cannot extend the spacetime beyond horizon, which therefore forms a natural boundary of spacetime. Similar singularities were studied in the context of cosmological models with homogeneous spatial section and it was referred to as the ``wimper singularity'' in that the universe terminates at a singularity where all physical quantities are well-behaved (a ``whimper'' rather than a ``bang'') and an associated Cauchy horizon \cite{Ellis:1974ug,siklos}.

    It should be emphasized that having an irrational $\Delta_+$ is not a consequence of
higher-derivative corrections. This is because the dynamics of perturbative equation is governed by the linearized equation of Einstein gravity in the background of the rotating black hole. Our conclusion therefore applies to Einstein gravity with minimally coupled matter, unless the matter Lagrangian is so fine tuned that these irrational terms all drop out. The reason that we focus on higher-derivative corrections in this paper is to restrict our tension to only pure gravities, which however are not necessary to demonstrate the vulnerability of the horizon regularity.

The paper is organized as follows. In section 2, we review the Reall-Santos procedure to obtain the corrected black hole thermodynamic variables without solving for the perturbative solutions. This allows us to double check our later numerical calculations of the perturbative solutions. In section 3, we construct linear perturbation of the $D=7$ rotating black hole with all equal angular momenta in Einstein gravity extended by a Riemann-squared term. We analyse both the horizon and asymptotic structure and obtain the numerical solution that validates the result in section 2. We generalize the $D=7$ discussion to general $D=2n+1$ dimensions in section 4. We then study the implication of having irrational powers in the near-horizon structure in section 5. We conclude our paper in section 6. In appendices A, B and C, we present some detailed complicated formulae that would interrupt the discussion if presented in the main text.

\section{Corrections to Thermodynamics}
\label{sec:thermo}

In this paper, we focus on the study of rotating black holes with all equal angular momenta in general $D=2n+1$ dimensions. We shall restrict ourselves to pure gravity for simplicity. Many exact solutions of Ricci-flat or Einstein metrics have been constructed, {\it e.g.},
\cite{Kerr:1963ud, Carter:1968ks, Myers:1986un,Hawking:1998kw,Gibbons:2004js,Gibbons:2004uw, Chen:2006xh,Lu:2008js}.  The most general quadratic curvature extension to Einstein gravity
\be
S_{\rm Ein} = \fft{1}{16\pi} \int d^D x\,\sqrt{-g}\, R\,,
\ee
involves three terms, {\it i.e.}~$R^{\mu\nu\rho\sigma}R_{\mu\nu\rho\sigma}$, $R^{\mu\nu}R_{\mu\nu}$ and $R^2$. However, in the effective field theory approach, the couplings of these terms are all small, and hence the latter two terms can be removed by the field redefinition $g_{\mu\nu}\rightarrow g_{\mu\nu} + c_1 R_{\mu\nu} + c_2 R g_{\mu\nu}$, leading to an equivalent description of quadratic curvature correction with simply just one term
\be
S_{\rm quad} = \fft{1}{16\pi} \int d^D x\, \sqrt{-g} (\alpha\,R^{\mu\nu\rho\sigma}R_{\mu\nu\rho\sigma})\,.\label{quadraticlag}
\ee

Rotating black holes that are asymptotically flat in general dimensions were constructed in \cite{Myers:1986un}. In odd $D=2n+1$ dimensions, when all the angular momenta are equal, the metric reduces to cohomogeneity-one depending only on the radial variable $r$. Following the notation in \cite{Feng:2016dbw}, this solution can be expressed as
\bea
ds^2_{2n+1}=-\frac{h(r)}{W(r)}dt^2+\frac{dr^2}{f(r)}+r^2W(r)(\sigma+\omega(r)dt)^2
+r^2ds^2_{\mathbb{CP}^{n-1}}\,.\label{rotating-metric}
\eea
Here $ds^2_{\mathbb{CP}^{n-1}}$ is the metric of an $(n-1)$-dimensional complex projective space. (We adopt the convention of \cite{Cvetic:2018ipt} for the $\mathbb{CP}^n$ metrics.)  The metric functions are given by $h=f=\bar f$, $W=\bar W$ and $\omega=\bar \omega$, with
\bea
\bar W=1+\frac{\nu^2}{r^{D-1}},\quad \bar \omega(r)=\frac{\sqrt{\mu}\nu}{r^{D-1} \bar W},\quad \bar f=1-\frac{\mu}{r^{D-3}} +\frac{\nu^2}{r^{D-1}}\,.\label{Einstein-solution}
\eea
The solution has two integration constants $(\mu,\nu)$, parameterizing the mass $M_0$ and the angular momentum $J_0$. The event horizon $r_{0}$ is located at the largest root of $\bar f$ and thermodynamical quantities can all be easily obtained, given by
\bea
M_0&=&\frac{  (D-2) \Omega _{D-2}}{16 \pi }\mu,\qquad J_0=\frac{(D-1) \Omega _{D-2} }{16 \pi }\sqrt{\mu } \nu ,
\qquad \Omega_0=\frac{\nu }{r_{0} \sqrt{r_{0}^{D-1}+\nu ^2}},\cr
%%%
T_0&=&\frac{(D-3) r_{0}^{D-1}-2  \nu ^2}{4 \pi  r_{0}^{\frac{1}{2}(D+1)} \sqrt{r_{0}^{D-1}+\nu ^2}},
\qquad S_0=\frac{\Omega _{D-2}}{4}  r_{0}^{\frac{1}{2}(D-3)} \sqrt{r_{0}^{D-1}+\nu ^2},
\eea
where $\Omega_{D-2} \equiv \fft{(D-1)}{\Gamma((D+1)/2)} \pi^{(D-1)/2}$ is the volume of the unit round $(D-2)$-sphere. The parameters $(\mu,\nu)$ and the horizon location $r_{0}$ are related by $\bar f(r_{0})=0$. The Gibbs free energy associated with the Euclidean action is
\bea
G_0=M_0-T_0S_0-\Omega_0J_0\,.
\eea
The leading $\alpha$-order  correction from the quadratic extension \eqref{quadraticlag} to these thermodynamic quantities can be computed using the Reall-Santos method \cite{Reall:2019sah} without having to construct the corresponding corrected solution. Specifically, we need first to fix the temperature and angular velocity
\bea
T=T_0+\mathcal{O}(\alpha^2)\,,\qquad \Omega=\Omega_0+\mathcal{O}(\alpha^2)\,.
\eea
The Gibbs free energy is then shifted by the correction from the quadratic extension:
\bea
G(T_0,\Omega_0,\alpha)&=&G_0+\Delta G+\mathcal{O}(\alpha^2)\,.
\eea
This shifted term can be calculated from the quadratic Euclidean action evaluated on the leading-order solution \eqref{Einstein-solution}. We find
\bea
\Delta G=\frac{\Delta I}{\beta}\,,\qquad \Delta I=-\frac{1}{16\pi}\int_0^\beta d\tau
\int_0^{r_{0}} dr\int d\Omega_{D-2}\sqrt{g} (\alpha R^{\mu\nu\rho\sigma} R_{\mu\nu\rho\sigma})\,.
\eea
We therefore have
\bea
\Delta G=-\frac{(D-3) \Omega _{D-2}}{16 \pi
   r_{0}^{D+3}}\left((D-2)^2 r_{0}^{2 (D-1)}-2 (2 D-3) r_{0}^{D-1} \nu ^2+\nu ^4\right)\alpha.
\eea
This allows one to obtain the complete set of thermodynamic variables at the $\alpha$-order correction, namely
\be
J=-\left(\frac{\partial G}{\partial\Omega_0}\right)\Big|_{T_0,\alpha},\qquad S=-\left(\frac{\partial G}{\partial T_0}\right)\Big|_{\Omega_0,\alpha},\qquad
M= G+T_0S+\Omega_0J\,.
\ee
In the extremal limit, $f(r_0)=0=f'(r_0)$, {\it i.e.},
\be
\mu = \ft12(D-1) r_0^{D-3}\,,\qquad \nu^2=\ft12(D-3) r_0^{D-1}\,,\label{extcondgend}
\ee
the mass and angular momentum are no longer independent, but satisfies the relation
\bea
M_{\mathrm{ext}}
&=&\frac{ (D-2) \left(\frac{\Omega _{D-2}}{32 \pi }\right){}^{\frac{1}{D-2}} }{(D-3)^{\frac{D-3}{2 (D-2)}} (D-1)^{\frac{D-5}{2
   (D-2)}}}\, J_{\mathrm{ext}}^{\frac{D-3}{D-2}}\cr
   &&+\frac{(D-3)^{\frac{D+1}{2 (D-2)}} (3 D-11)\left(\frac{\Omega _{D-2}}{32 \pi }\right){}^{\frac{3}{D-2}}}{2 (D-1)^{\frac{D-11}{2 (D-2)}}}\, \alpha J_{\mathrm{ext}}^{\frac{D-5}{D-2}}\,.
\eea
The $D=5$ result was obtained in \cite{Ma:2020xwi}.  This is an elegant approach that makes finding the perturbative solution unnecessary. However, this procedure assumes that such a solution necessarily exists, which may not be guaranteed. Even if such a solution does exist, the procedure only uses the information of the spacetime regions outside the horizon. It does not tell how the inside of the horizon changes under the perturbation and how the geodesics can be extended on the horizon.

\section{Leading-order correction: the $D=7$ example}
\label{sec:D=7}

We now consider the leading $\alpha$-order correction to Ricci-flat rotating black holes \eqref{rotating-metric} while keeping the full isometry. The perturbative ans\"atze of the metric functions are chosen to be
\begin{eqnarray}
&W(r)=\bar W+\alpha \delta W,\qquad \omega(r)=\bar \omega - \alpha \fft{\bar \omega}{\bar W} \delta W+ \alpha \delta \omega\,,\nn\\
&f(r)=\bar f\,(1+\alpha \delta f),\qquad h(r)=\bar f\,(1+\alpha \delta h)\,.
\label{leadingorder}
\end{eqnarray}	
The linear perturbative equations of $(\delta f, \delta h, \delta W, \delta \omega)$ are not solvable analytically in general higher dimensions. We therefore choose to solve the perturbative equations with the horizon radius $r_0$ fixed for an easier numerical approach. In this set of ans\"atze, the perturbative functions $(\delta f, \delta h, \delta W, \delta \omega)$ should be all finite in all the regions from the outer horizon to asymptotic infinity. The $D=5$ case can be solved exactly, and was obtained and analyzed in \cite{Ma:2020xwi}.  However, analytic solutions for general $D$ do not seem to exist, and we focus on the $D=7$ case in this section and summarize the main results for higher dimensions in the next section.

\subsection{Perturbative equations}

The four independent coupled linear equations of $(\delta f, \delta h, \delta W, \delta \omega)$ in general dimensions were given in the appendix \ref{app:eom}. By the standard procedure of eliminating variables, we obtain a fourth order linear differential equation with a source for $\delta f$:
\be
P_4\, \delta f'''' + P_3\, \delta f''' + P_2\, \delta f'' + P_1\, \delta f' + P_0\, \delta f= Q\,.\label{deltafgeneq}
\ee
In $D=7$ dimensions, we have
\bea
P_4 &=& 5 r^{10} \left(\nu ^2+r^6-\mu  r^2\right)^2 \left(8 \mu  \nu ^2+27 r^{10}-10 \mu  r^6-5 \mu ^2 r^2\right),\cr
P_3 &=& 10 r^{9} \left(\nu ^2+r^6-\mu  r^2\right)(-32 \mu  \nu ^4+324 r^{16}-86 \mu  r^{12}-243 \nu ^2 r^{10}-80 \mu ^2 r^8\cr
&& +206 \mu  \nu ^2 r^6+10 \mu ^3 r^4+\mu ^2 \nu ^2 r^2),\cr
P_2 &=& 5 r^8 \Big(168 \mu  \nu ^6+3915 r^{22}-1848 \mu  r^{18}-2376 \nu ^2 r^{16}-974 \mu ^2 r^{14}+1584 \mu  \nu ^2 r^{12}\cr
&&+3 r^{10} \left(200 \mu ^3+819 \nu ^4\right)-712 \mu ^2 \nu ^2 r^8+r^6 \left(35 \mu ^4-774 \mu  \nu ^4\right)-96 \mu ^3 \nu ^2 r^4+\mu ^2 \nu ^4 r^2\Big),\cr
P_1 &=& 5 r^7 \Big(-168 \mu  \nu ^6+2565 r^{22}+504 \mu  r^{18}-2808 \nu ^2 r^{16}-466 \mu ^2 r^{14}+1776 \mu  \nu ^2 r^{12}\cr
&&-7 r^{10} \left(40 \mu ^3+351 \nu ^4\right)+136 \mu ^2 \nu ^2 r^8+3 r^6 \left(15 \mu ^4+386 \mu  \nu ^4\right)-64 \mu ^3 \nu ^2 r^4-\mu ^2 \nu ^4 r^2\Big),\cr
P_0 &=& -480 r^{12} \left(8 \mu  \nu ^4+135 r^{16}+14 \mu  r^{12}-108 \nu ^2 r^{10}-25 \mu ^2 r^8+40 \mu  \nu ^2 r^6-4 \mu ^2 \nu ^2 r^2\right),
\eea
together with the source
\bea
Q &=& 384 \Big(8960 \mu ^2 \nu ^6+1782 \mu  \nu ^2 r^{16}+135 r^{14} \left(101 \mu ^3+96 \nu ^4\right)-62340 \mu ^2 \nu ^2 r^{12}\cr
&&+r^{10} \left(54720 \mu  \nu ^4-5250 \mu ^4\right)+23610 \mu ^3 \nu ^2 r^8-r^6 \left(1125 \mu ^5+24352 \mu ^2 \nu ^4\right)\cr
&&+60 r^4 \left(121 \mu ^4 \nu ^2+24 \mu  \nu ^6\right)-14080 \mu ^3 \nu ^4 r^2\Big).
\eea
It is clear that the left-hand side of the equation, the linear sector, comes
from the linearization of the Einstein tensor $\delta G_{\mu\nu}$ on the rotating black hole background \eqref{Einstein-solution}. The right-hand side of the equation, the source $Q$, comes  from the contributions of $R^{\mu\nu\rho\sigma} R_{\mu\nu\rho\sigma}$, but evaluated on the original metric functions \eqref{Einstein-solution}.

For the remaining perturbations, $(\delta h, \delta \omega)$ reduce to quadratures and $\delta W$ becomes purely algebraic:
\bea
&&\delta h'(r)=-\frac{1}{5 r^9 \big(8 \mu  \nu ^2+27 r^{10}-10 \mu  r^6-5 \mu ^2 r^2\big)}\Big(96 \big(-416 \mu  \nu ^4+135 \mu ^2 r^8-312 \mu  \nu ^2 r^6\cr
&&-15 r^4 \big(11 \mu ^3+8 \nu ^4\big)+588 \mu ^2 \nu ^2 r^2\big)+120 r^{12}  \big(-2 \nu ^2+r^6+\mu  r^2\big)\delta f(r)\cr
&&+5 r^7  \big(7 \nu ^4+19 r^{12}-8 \mu  r^8+17 \nu ^2 r^6-7 \mu ^2 r^4+\mu  \nu ^2 r^2\big)\delta f'(r)\nn\\
&&+35 r^8 \big(2 r^6-\nu ^2\big)  \big(\nu ^2+r^6-\mu  r^2\big)\delta f''(r)+5 r^9  \big(\nu ^2+r^6-\mu  r^2\big)^2\delta f^{(3)}(r)\Big),\\
%%%%%%
&&\delta W =-\frac{36 \mu  \big(5 \mu  r^2-7 \nu ^2\big)}{5 r^{12}}+\big(\frac{\nu ^2}{r^6}-2\big)\delta f(r)-\frac{ \big(\nu ^2+r^6-\mu  r^2\big)}{4 r^5}\big(\delta f'(r)+\delta h'(r)\big),\\
%%%%%
&& \delta\omega'(r)=\frac{1}{6 \sqrt{\mu } \nu  r^{11} \big(\nu ^2+r^6\big)^2}\Big(24 \big(\nu ^2+r^6\big)^2 \big(16 \nu ^4+15 \mu ^2 r^4-36 \mu  \nu ^2 r^2\big)+4 r^8  \big(5 r^6-\nu ^2\big)\cr
&& \big(\nu ^2+r^6\big)^2\delta f(r)-18 \mu  \nu ^2 r^{16}\delta h(r) +r^9 \big(2 \nu ^4+5 r^{12}+7 \nu ^2 r^6\big)  \big(\nu ^2+r^6-\mu  r^2\big)\delta h'(r)\\
		&&+2 r^{14}  \big(8 \nu ^4+2 r^{12}+\nu ^2 r^6+9 \mu  \nu ^2 r^2\big)\delta W(r)-r^{15} \big(\nu ^2+r^6\big)  \big(-2 \nu ^2+r^6-3 \mu  r^2\big)\delta W'(r)\Big).\nn
\eea
These equations are sufficiently complicated that we need to apply numerical methods to connect the near horizon geometry to the asymptotic infinity, both of which can be analysed analytically.

\subsection{Asymptotic behavior}

Although the peturbative equations cannot be solved exactly, we can nevertheless obtain the asymptotic behavior. Assuming that the leading-order behavior is $\delta f = r^\gamma f_0$, we have
\be
(\gamma-2)(\gamma+4)(\gamma+6)(\gamma+10)\, r^{\gamma+12}\,f_0 =- \ft{25344}5\mu\nu^2\,.
\ee
Therefore, we have $\gamma=2,-4,-6,-10$ for the source-free contributions and $\gamma=-12$ for the source contribution, leading to the general solution with four integration constants
\be
\delta f=-\frac{132 \mu  \nu ^2}{35 r^{12}} \tilde f_0(r) +\frac{c_{10}}{r^{10}}\tilde f_{10}(r) +\frac{c_{6}}{r^6}\tilde f_{6}(r)+\frac{c_4}{r^4}\tilde f_4(r)+ c_{-2}\, r^2 \tilde f_{-2}(r)\,,\label{largerdfstr}
\ee
where $\tilde f_i$'s all take the form of $\tilde f_i \sim 1 + \#_1/r + \#_2/r^2 + \cdots$ at the large $r$ expansion. (We use $\#_i$ to denote some generic constants.)

The first term in \eqref{largerdfstr} comes from the quadratic curvature extension. It has faster falloff than any other terms, which come from the perturbation of the Einstein tensor. The $c_{-2}$ term is rather intriguing since it appears like a cosmological constant in the $g_{rr}$ metric component.  This term however will not arise from the linear perturbation equation of the Schwarzschild black hole, while keeping the static and spherical symmetry. The connection between rotation and cosmological constant is worth further exploring.

In our case, however, in order for $\delta f$ to be regular at asymptotic infinity, we must set $c_{-2}=0$. Therefore, general asymptotically-flat perturbations involve three independent free parameters parameters $(c_4,c_6,c_{10})$. The low-lying falloff orders of $\tilde f_i$, so that $\delta f$ is up to order $1/r^{16}$, are given by
\bea
&&\tilde f_0=1-\frac{21 \mu }{11 r^4}+\cdots\,,\qquad
\tilde f_4= 1+\frac{\mu }{r^4}+\frac{\mu ^2}{r^8}-\frac{59 \mu  \nu ^2}{160 r^{10}} + \frac{\mu ^3-\frac{2 \nu ^4}{5}}{r^{12}}\cdots\,,\cr
&&\tilde f_6=1-\frac{\nu ^2}{r^6}-\frac{101 \mu ^2}{160 r^8} -\frac{3 \mu  \nu ^2}{5 r^{10}}+ \cdots \,,\qquad \tilde f_{10}=1 + \frac{261 \mu }{160 r^4} -\frac{7 \nu ^2}{5 r^6} + \cdots\,.
\eea
Analogously, we can obtain the leading falloffs of the $(\delta h, \delta W, \delta\omega)$ functions
\bea
\delta h &=&  \Big(\frac{48 \mu ^2}{5 r^{10}}-\frac{564 \mu  \nu ^2}{35 r^{12}}+
\frac{3 \left(693 \mu ^3+608 \nu ^4\right)}{280 r^{14}}-\frac{744 \mu ^2 \nu ^2}{35 r^{16}} + \cdots\Big) \cr
&& + \fft{c_{10}}{5r^{10}}\Big(1+\frac{61 \mu }{32 r^4}-\frac{2 \nu ^2}{r^6} + \cdots) +
\fft{c_6}{r^6}\Big(1+\frac{4 \mu }{5 r^4}+\frac{99 \mu ^2}{160 r^8}-\frac{\nu ^2}{r^6}-\frac{8 \mu  \nu ^2}{5 r^{10}}+ \cdots\Big)\cr
&&+\fft{c_4}{r^{4}}\Big(1+\frac{\mu }{r^4}+\frac{\mu ^2}{r^8}+\frac{\mu ^3}{r^{12}}-\frac{4 \nu ^2}{5 r^6}-\frac{259 \mu  \nu ^2}{160 r^{10}}+\frac{3 \nu ^4}{5 r^{12}}+\cdots\Big),\\
\delta W &=& \Big(-\frac{12 \mu ^2}{r^{10}}-\frac{12 \mu  \nu ^2}{7 r^{12}}+\frac{456 \mu ^2 \nu ^2}{35 r^{16}}-\frac{3 \left(25 \mu ^3-32 \nu ^4\right)}{8 r^{14}} + \cdots\Big)\cr
&& + \fft{c_{10}}{r^{10}} \Big(1+\frac{25 \mu }{32 r^4}-\frac{2 \nu ^2}{5 r^6} + \cdots\Big)
+ \fft{c_6}{r^6} \Big( 1-\frac{\mu }{r^4}-\frac{25 \mu ^2}{32 r^8}+\frac{2 \mu  \nu ^2}{5 r^{10}} + \cdots \Big)\cr
&& + \fft{c_4\nu^2}{r^{10}} \Big(1+\frac{25 \mu }{32 r^4}-\frac{2 \nu ^2}{5 r^6}+ \cdots\Big),\\
\delta\omega &=& \frac{8 \mu ^{3/2}}{\nu  r^6} \Big(1-\frac{5 \nu ^2}{2 r^6}-\frac{3 \left(7105 \mu ^3+3312 \nu ^4\right)}{15680 \mu r^8}+\frac{\nu ^2 \left(18027 \mu ^3+4256 \nu ^4\right)}{4480\mu ^2 r^{10}} + \cdots\Big)\cr
&&-\frac{2c_{10}}{3 \sqrt{\mu } \nu  r^6} \Big(1-\frac{7 \mu ^2}{64 r^8}-\frac{\nu ^2}{r^6}-\frac{291 \mu  \nu ^2}{640 r^{10}} + \cdots\Big)+\frac{7 \sqrt{\mu }\, c_{6}}{6 \nu  r^6} \Big(1-\frac{\nu ^2}{r^6}+\frac{2109 \mu  \nu ^2}{1120 r^{10}}\cr
&&-\frac{87 \mu ^3+80 \nu ^4}{112 r^8 \mu } + \cdots\Big)-\frac{7 \nu\,c_4 }{6 \sqrt{\mu } r^6}\Big(1-\frac{247 \mu ^2}{112 r^8}+\frac{5 \mu ^4}{7 r^{10} \nu ^2}-\frac{\nu ^2}{r^6}+\frac{2109 \mu  \nu ^2}{1120 r^{10}}+ \cdots\Big).
\eea
From these asymptotic behavior, we can read off the mass $M$ and angular momentum $J$, at their their $\alpha$-order correction:
\be
M=\frac{5\pi^2}{16}(\mu-\alpha c_4) \,,\qquad
J=\frac{3\pi^2}{8}\sqrt{\mu}\nu \Big(1 + \fft{\alpha}{6 \mu  \nu ^2}(48 \mu ^2-7 \nu ^2 c_4+7 \mu c_6-4 c_{10})\Big)\,.\label{mjpert}
\ee

\subsection{Near-horizon structure}
\label{sec:sub-nhne}

Note that in our perturbative solution, we hold the horizon radius $r_0$ fixed; therefore, both the Mass and angular momentum acquire higher-order corrections. At the first sight, the perturbed mass and angular momentum \eqref{mjpert} depends on three parameters $(c_4,c_6,c_{10})$, which would violate the no-hair theorem, since we would expect that the solution should only have two independent parameters. Indeed not all the parameters $(c_4,c_6,c_{10})$ leads to black hole solutions.

In order to determine the parameter choices of $(c_4,c_6,c_{10})$ that yield black holes, we can study the near-horizon geometry. For general non-extremal black holes, the analysis will be given in appendix \ref{app:nhne}.  Here, we shall focus only on the extremal case, corresponding to
\be
\mu=3 r_0^4\,, \qquad \nu=\sqrt{2}r_0^3\,.\label{extcond}
\ee
In other words, the solution depends only on one parameter, such that mass and angular momentum are no longer independent, but they satisfy
\be
M_0= \frac{5 \pi ^{2/5}}{4 \sqrt[5]{3}}\, J_0^{\fft45}\,.\label{mjleading}
\ee
To determine the leading-order behavior of $\delta f$ as $r\rightarrow r_0$, we can adopt the same trick by assuming $\delta f=(r-r_0)^\gamma \hat f$, where $\tilde f$  is analytic, satisfying the usual Taylor expansion at $r=r_0$.  We find that the leading-order equation for small $(r-r_0)$ is
\be
(\gamma+1)(\gamma+2)(\gamma - \Delta_+)(\gamma - \Delta_-)\, r^{-\gamma}\,\hat f(r_0) = - \fft{168}{5 r_0^2}\,,
\ee
where
\be
\Delta_\pm =-\frac{3}{2}\pm \frac{\sqrt{21}}{2}\,.\label{d7delta}
\ee
Thus the general solution near $r_0$ takes the form
\be
\delta f = \fft{28}{5 r_0^2}\hat f_0 + \fft{d_{-1}}{r-r_0} \hat f_{-1} + \fft{d_{-2}}{(r-r_0)^2}\hat f_{-2} + d_{\Delta_-} (r-r_0)^{\Delta_-} \hat f_{\Delta_-} + d_{\Delta_+} (r-r_0)^{\Delta_+}\hat
f_{\Delta_+}\,.
\ee
Note that $\Delta_+>0$ and $\Delta_-<0$. The regularity at $r_0$ requires that we set coefficients $d_{-1}, d_{-2}$ and $d_{\Delta_-}$ all to zero.  The low-lying orders of the near-horizon expansions at $r=r_0$ are thus given by
\bea
\delta f &=& \fft{28}{5 r_0^2}\Big(1 +\frac{3658}{21 r_0}\left(r-r_0\right)-\frac{63167 }{441 r_0^2}\left(r-r_0\right)^2+\cdots\Big)+ d_{\Delta_+} (r-r_0)^{\Delta_+} \bigg(1\nn\\
&&+ \frac{117-7 \sqrt{21}}{180 r_0}\left(r-r_0\right)  -\frac{1282 \sqrt{21}-6627}{18360 r_0^2}\left(r-r_0\right)^2 + \cdots \bigg).\label{dfhorizon}
\eea
Analogous solutions can be obtained for $(\delta h, \delta W, \delta\omega)$, which we present in appendix \ref{app:nhe}.  The irrational power of $(r-r_0)^{\Delta_+}$ should cause our concern since it indicates that the near-horizon geometry is not analytic.  We shall come back to this point in section \ref{sec:boundary}.

We now continue to focus on the $\delta f$ equation. On the horizon, the general regular solution of $\delta f$ is specified by two parameters, namely $r_0$ and coefficient $d_{\Delta_+}$. However, for a given $r_0$, a generic value of $d_{\Delta_+}$ will excite the coefficient $c_{-2}$ of $r^2$ in the large-$r$ expansion \eqref{largerdfstr}.  We need to fine-tune the value $d_{\Delta_+}$ for each given $r_0$ so that $c_2$ vanishes and the metric remains asymptotically flat. If such a $d_{\Delta_+}$ exists, then we obtain a leading-order perturbation the extremal rotating black hole. Thus, in the linearly perturbed solution, both the horizon and asymptotic parameters $(d_{\Delta_+}, c_4, c_6, c_{10})$ all depend on the horizon radius $r_0$, analogous to the unperturbed extremal black hole, which has only one independent parameter. This is consistent with the no-hair theorem. Consequently, it follows from \eqref{mjpert} that both mass and angular momentum depend only on $r_0$:
\be
M=\ft{15}{16} \pi^2 r_0^4 + \alpha \delta M(r_0)\,,\qquad
J=\ft34 \sqrt{\ft32} \pi^2 r_0^5 + \alpha \delta J(r_0)\,.
\ee
Therefore, at the linear $\alpha$ order, the mass-angular momentum relation becomes
\be
M=\frac{5 \pi ^{2/5}}{4 \sqrt[5]{3}} J^{\fft45}\Big(1 +\eta \alpha J^{-\fft25}\Big),
\ee
where $\eta$ is some order-one dimensionless constant.  Analytical calculation using the Reall-Santos procedure discussed in section \ref{sec:thermo} indicates that
\be
\eta = 2\ 3^{3/5} \pi ^{4/5} \sim 9.661\,.\label{d7eta}
\ee
From the leading-order perturbative solution, we deduce
\be
\eta=\frac{\pi ^{4/5}}{30\ 3^{2/5}}\Big(-\frac{c_4}{r_0^2}-\frac{21 c_6}{r_0^4}+\frac{4 c_{10}}{r_0^8}-432\Big).
\ee
We use numerical approach to show that the perturbative solution outside the horizon correctly reproduce the value of the $\eta$ coefficient \eqref{d7eta}.

\subsection{Numerical result}

Since we do not have analytical solution of $\delta f$, we adopt the numerical approach to determine the coefficient $d_{\Delta_+}$ so that $c_2$ vanishes. Specifically, we use the shooting method by beginning with the analytical result of the power series expansion \eqref{dfhorizon} up to $(r-r_0)^4$ order as the initial data, and then integrate over to large $r$, {\it e.g.}~$200r_0$. We search for the appropriate $d_{\Delta_+}$ to shoot the target of $\delta f\rightarrow 0$.  We then do curvature fitting against the asymptotic structure \eqref{largerdfstr} up to the order $1/r^{14}$. We first find a fine-tuned result of  $d_{\Delta_+}$ so that $c_{-2}$ vanishes. We then drop off the $c_{-2}$ term and do curvature fitting against the remaining coefficients and read off $c_{4}, c_6$ and $c_{10}$.
This allows us to obtain the $\alpha$-order correction to the mass-angular momentum relation.

\begin{figure}[ht!]
\begin{center}
\includegraphics[width=440pt]{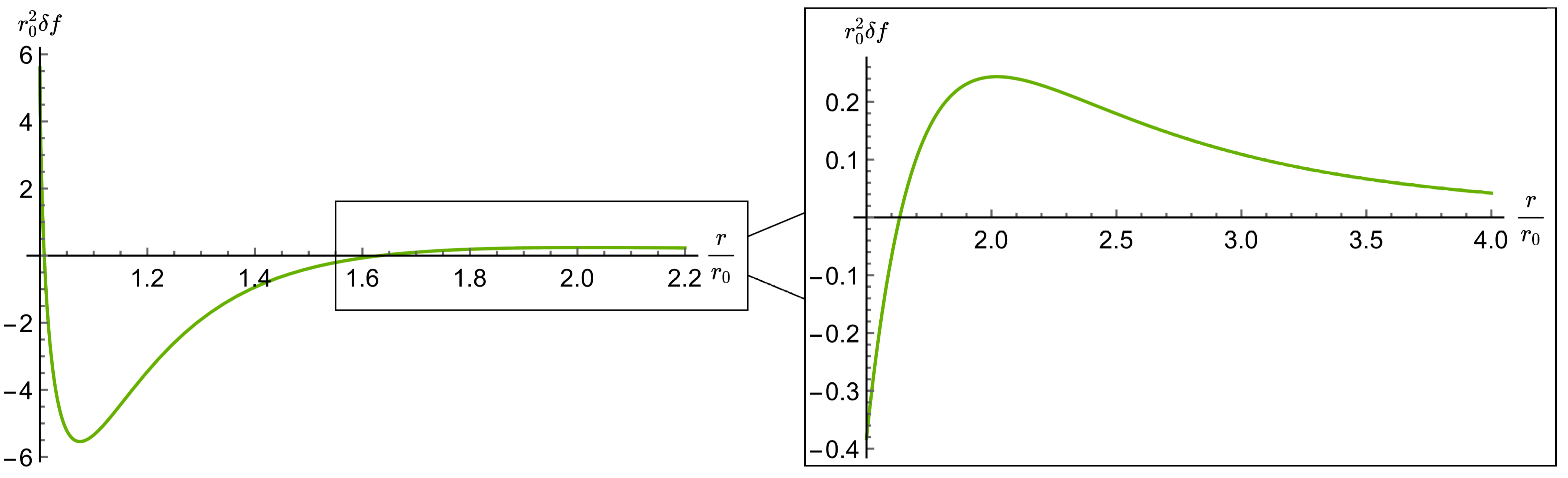}
\end{center}
\caption{{\it\small The dimensionless function $r_0^2\delta f$ is not a monotonous function of the dimensionless radial variable $\tilde r=r_0/r$. The numerical result allows us to read off the asymptotic falloffs and determine both corrected mass and angular momentum.}}
\label{deltaffig}
\end{figure}

The result depends on $r_0$ trivially, since it is the only scale parameter in the linearized equation. To see this explicitly, we can define a dimensionless radial coordinate $\tilde r=r/r_0$, in which case, $r_0$ drops out from the linear equation completely for the dimensionless function $(r_0^2 \delta f)$. We can therefore perform the numerical analysis for $r_0=1$ without loss of generality. We plot dependence of the dimensionless function $r_0^2\delta f$ on the dimensionless radial variable $\tilde r$ in Fig.~\ref{deltaffig}. We see that $r_0^2\delta f$ is not a monotonous function of $\tilde r$.  From this numerical solution, we can read off the asymptotic structure and determine both corrected mass and angular momentum. We obtain the expected $\eta$ within $1\%$ of accuracy.  We summarize the result in Table 1.

\begin{center}
\begin{tabular}{|c|c|c|c|c|c|c|c|}
  \hline
  % after \\: \hline or \cline{col1-col2} \cline{col3-col4} ...
 $d_{\Delta_+} r_0^{2+\Delta_+}$  & $c_4/r_0^2$ & $c_6/r_0^4$ & $c_{10}/r_0^8$ &$\delta M/r_0^2$ &$\delta J/r_0^3$ & $\eta$\\ \hline
  -604.856  & 13.440  & -45.165 & -80.72 &-41.452 &-96.133 & 9.669\\
\hline
\end{tabular}
\bigskip

{\small Table 1. Numerical data of the $\alpha$-correction of the extremal rotating black hole.}
\end{center}

\section{General $D=2n+1$ dimensions}
\label{sec:gend}

In the previous section, we found that in $D=7$ dimensions, the near-horizon geometry of the extremal rotating black hole involves $(r-r_0)^\Delta_+$ terms with irrational $\Delta_+$. Such an irrational power does not arise in $D=5$. In this section, we study whether such an irrational power occurs in general higher dimensions or it is a special case in $D=7$.  The linear perturbative equations in general $D=2n+1$ dimensions are given in appendix \ref{app:eom}. By eliminating variables, we obtain the decoupled $\delta f$ equation, as in the $D=7$ case. It is given by
\be
P_4 \delta f'''' + P_3 \delta f''' + P_2 \delta f'' + P_1 \delta f' + P_0 \delta f= Q\,,
\ee
where $P_i$ and $Q$ are polynomials of $r$. We find
\bea
&&P_4 = (D-2) r^{D+4} \left(r^D-\mu  r^3+\nu ^2 r\right)^2 \Big(-8 (D-2) \mu  r^{D+3}+3 (D-1)^2 r^{2 D}-4 (D-2) \mu ^2 r^6\cr
&&\,\, +4 (D+1) \mu  \nu ^2 r^4\Big),\cr
%%%%%
%%%%%
&&P_3 = (D-2) r^{D+3} \left(r^D-\mu  r^3+\nu ^2 r\right)\Big(-8 \left(5 D^2-29 D+38\right) \mu ^2 r^{D+6}\cr
&&\,\,+8 \left(5 D^2-5 D-4\right) \mu  \nu ^2 r^{D+4}-4 \left(D^2-13 D+22\right) \mu ^3 r^9+4 \left(2 D^2-11 D-19\right) \mu ^2 \nu ^2 r^7\cr
&&\,\,+\left(9 D^3-109 D^2+247 D-163\right) \mu  r^{2 D+3}-9 (D-1)^3 \nu ^2 r^{2 D+1}+12 (D-1)^3 r^{3 D}\cr
&&\,\,-4 (D+1)^2 \mu  \nu ^4 r^5\Big),\nn\\
%%%%%%
%%%%%
&&P_2 = (D-2) r^{D+2}\Big(
-3 (D-1)^2 \left(2 D^2-D-3\right) \nu ^2 r^{3 D+1}+3 (D-1)^2 \left(5 D^2-16 D+12\right) r^{4 D}\cr
&&\,\,+4 \left(5 D^2-38 D+56\right) \mu ^4 r^{12}+4 \left(-7 D^2+27 D+58\right) \mu ^3 \nu ^2 r^{10}-4 \left(D^2-8 D+6\right) \mu ^2 \nu ^4 r^8\cr
&&\,\,-4 \left(D^3-43 D^2+200 D-236\right) \mu ^3 r^{D+9}+4 \left(2 D^3-32 D^2+23 D+9\right) \mu ^2 \nu ^2 r^{D+7}\cr
&&\,\,-4 \left(D^3+11 D^2-13 D-17\right) \mu  \nu ^4 r^{D+5}-2 \left(6 D^4-73 D^3+157 D^2-17 D-109\right) \mu  \nu ^2 r^{2 D+4}\cr
&&\,\,+3 \left(2 D^4-37 D^3+139 D^2-215 D+119\right) \mu  r^{3 D+3}+3 (D-1)^2 D (2 D-1) \nu ^4 r^{2 D+2}\cr
&&\,\,+ \mu  r^6 \left(\left(6 D^4-131 D^3+702 D^2-1199 D+626\right) \mu  r^{2 D}+12 D (D+1) \nu ^6\right)\Big),\nn\\
%%%%%%
%%%%%%
&& P_1= (D-2) r^{D+1}\Big(12 \left(D^2-6 D+8\right) \mu ^4 r^{12}-4 \left(D^2+11 D-62\right) \mu ^3 \nu ^2 r^{10}\cr
&&\,\,+4 \left(D^2-8 D+6\right) \mu ^2 \nu ^4 r^8-3 (D-1)^2 \left(D^3-5 D^2+6\right) \nu ^2 r^{3 D+1}+3 (D-1)^2 (2 D^3-17 D^2\cr
&&\,\,+38 D-24) r^{4 D}+4 \left(D^4-18 D^3+102 D^2-245 D+210\right) \mu ^3 r^{D+9}-4(2 D^4-25 D^3\cr
&&\,\,+93 D^2-142 D+74) \mu ^2 \nu ^2 r^{D+7}-\left(6 D^4-67 D^3+278 D^2-495 D+282\right) \mu ^2 r^{2 D+6}\cr
&&\,\,+4 \left(D^4-7 D^3+25 D^2-5 D-32\right) \mu  \nu ^4 r^{D+5}+2(6 D^4-25 D^3-99 D^2+415 D\cr
&&-333) \mu  \nu ^2 r^{2 D+4}+\left(3 D^5-59 D^4+421 D^3-1267 D^2+1676 D-798\right) \mu  r^{3 D+3}\cr
&&\,\,-3 (D-1)^2 D (2 D-1) \nu ^4 r^{2 D+2}-12 D (D+1) \mu  \nu ^6 r^6
\Big),\cr
%%%%
%%%%
&&P_0 = -2 (D-3) (D-2) (D-1) r^{2 D}\Big(-4 \left(D^2-8 D+15\right) \mu ^2 \nu ^2 r^7+4 \left(D^2-4 D-5\right) \mu  \nu ^4 r^5\cr
&&\,\,+2 \left(D^3-16 D^2+57 D-58\right) \mu ^2 r^{D+6}-2 \left(D^3-10 D^2-9 D+50\right) \mu  \nu ^2 r^{D+4}+(5 D^3\cr
&&\,\,-51 D^2+147 D-133) \mu  r^{2 D+3}-3 (D-1)^2 (D+1) \nu ^2 r^{2 D+1}+6 (D-2) (D-1)^2 r^{3 D}\Big),
\eea
together with the source contribution
\bea
&&Q=2(D-1)\Big(-12 (D-1)^3 (D+1)^2 \left(D^2-5 D-24\right) \nu ^4 r^{3 D}-4 (D-3)^2 (D-2)^3 (D^2\cr
&&\,\,-5 D+4) \mu ^5 r^{13}-4 (D+1)^3 \left(D^4-42 D^2-64 D+105\right) \mu  \nu ^8 r^5-2 \left(D^2-5 D+6\right)^2(11 D^3\cr
&&\,\,-86 D^2+203 D-140) \mu ^4 r^{D+10}-3 (D-1)^3 (D+1)^2 \left(3 D^3-11 D^2-67 D-21\right) \nu ^6 r^{2 D+1}\cr
&&\,\,+6 (D-1)^3 \left(2 D^4-15 D^3+8 D^2+19 D-6\right) \mu  \nu ^2 r^{3 D+2}+2 (D+1)^2 (11 D^5-64 D^4\cr
&&\,\,-192 D^3+658 D^2+709 D-1122) \mu  \nu ^6 r^{D+4}+4 (D+1)^2 (4 D^5-16 D^4-91 D^3\cr
&&\,\,+107 D^2+279 D-315) \mu ^2 \nu ^6 r^7-8(3 D^7-21 D^6+112 D^4-127 D^3-23 D^2\cr
&&\,\,+156 D-36) \mu ^3 \nu ^4 r^9+3 (D-3)^2(3 D^6-38 D^5+194 D^4-516 D^3+759 D^2-586 D\cr
&&\,\,+184) \mu ^3 r^{2 D+7}+2(33 D^7-434 D^6+2090 D^5-4734 D^4+4987 D^3-1396 D^2-1350 D\cr
&&\,\,+804) \mu ^3 \nu ^2 r^{D+8}-2(33 D^7-280 D^6+311 D^5+1046 D^4-945 D^3-1988 D^2+729 D\cr
&&\,\,+1350) \mu ^2 \nu ^4 r^{D+6}+4 (4 D^7-48 D^6+201 D^5-341 D^4-53 D^3+1109 D^2-1580 D\cr
&&\,\,+732) \mu ^4 \nu ^2 r^{11}-(27 D^8-378 D^7+2101 D^6-6336 D^5+10905 D^4-10042 D^3+4139 D^2\cr
&&\,\,-524 D+108) \mu ^2 \nu ^2 r^{2 D+5}+(27 D^8-252 D^7+568 D^6-284 D^5-1562 D^4+1132 D^3\cr
&&\,\,+4192 D^2-596 D-3225) \mu  \nu ^4 r^{2 D+3}\Big).
\eea
In the extremal limit \eqref{extcondgend}, we can take an ansatz $\delta f = (r-r_0)^\gamma \hat f(r)$, where $\hat f(r)$ is analytic at $r=r_0$. In the limit of $r\rightarrow r_0$, we find that the leading-order equation is
\be
(\gamma + 1)(\gamma+2)(\gamma -\Delta_{-})(\gamma -\Delta_+) \hat f(r_0) = \frac{(D-9) (D-1) (D+7)}{(D-2) r_+^2}\,,
\ee
where
\be
\Delta_\pm =\fft12\left(-3\pm \sqrt{\frac{17 D-35}{D-3}}\right),
\ee
which reproduces the earlier $D=7$ result. Since only $\Delta_+>0$, the near-horizon structure of $\delta f$ takes the form
\be
\delta f = \frac{(9-D) (D-3) (D+7)}{4 (D-2) r_0^2} \hat f_0(r) + d_{\Delta_+}\, (r-r_0)^{\Delta_+}\, \hat f_{\Delta_+}(r),\label{dfgend}
\ee
where $d_{\Delta_+}$ a certain appropriate coefficient that is to be determined. Both functions $\hat f(r)$ and $ \hat f_{\Delta_+}(r)$ are analytic, satisfying the Taylor expansion of the form $1 + \#_1 (r-r_0) + \#_2 (r-r_0)^2 + \cdots$. Thus we see that in general higher dimensions, $\Delta_+$ is irrational as in the $D=7$ case. However, rational numbers can arise for sporadically distributed old $D$, and we give a few low-lying examples in Table 2.

\bigskip
\begin{center}
\begin{tabular}{|c|c|c|c|c|c|}
  \hline
  % after \\: \hline or \cline{col1-col2} \cline{col3-col4} ...
  $D$ & 5 & 101 & 1027 & 10661 & 428741 \\ \hline
  $\Delta_+$ & 1 & $\fft47$ & $\frac{9}{16}$  & $\frac{41}{73}$ & $\frac{260}{463}$ \\
  \hline
\end{tabular}
\bigskip

{\small Table 2. Sporadic dimensions that give rise to rational $\Delta_+$.}
\end{center}

It should be pointed out that there exists a decoupling limit $\epsilon\rightarrow 0$ associated with the coordinate transformation $r=r_0 + \epsilon \rho$. In this limit, the near-horizon geometry AdS$_3\times \mathbb{CP}^n$ becomes the solution on its own where AdS$_3$ is written as a $U(1)$ bundle over AdS$_2$. The $\Delta_+$ term drops out completely from the metric of the homogenous vacuum.

\section{Natural boundary on the horizon}
\label{sec:boundary}

We have found that the Riemann-squared correction to Einstein gravity has a consequence that the extremal rotating black holes in $D\ge 7$ odd dimensions with all equal angular momenta have in general irrational powers in the near-horizon expansion, as given in \eqref{dfgend}, where $\Delta_+$ is in general irrational. In this section we discuss its implication.

\subsection{Horizon as a natural boundary}

In a typical black hole, {\it e.g.}~the Schwarzschild black hole, the horizon ($g_{tt}=0$) represents a coordinate singularity that can be removed by coordinate extension such as the Kruskal extension. Such a procedure is problematic when the near-horizon geometry is not analytic with irrational powers.  A simpler toy model is the extremal RN black hole with certain correction
\be
ds^2 = - f dt^2 + \fft{dr^2}{f} + r^2 \Omega_2^2\,,\qquad
f=\Big(1-\fft{r_0}{r}\Big)^2 \Big(1 + \fft{\alpha}{r^3}\big(1-\fft{r_0}{r}\big)^{\Delta_+}\Big).
\ee
We require that $\alpha>0$ so that the metric satisfies both the strong and weak energy conditions on and outside the event horizon, {\it i.e.}~$r\ge r_0$. If $\Delta_+$ is a positive integer, then the function $f$ is analytic at $r=r_0$, and it is infinitely differentiable. However, in our case, $\Delta_+$ is an irrational positive number lying between 0 and 1. The function $f$ is therefore only 2'nd-order differentiable. However, the metric is still infinitely differentiable under covariant derivatives, owing to the fact that
\be
g^{rr} \nabla_r \nabla_r = f \partial_r^2 + \cdots\,.
\ee
Therefore, there is no curvature or covariant-derivative curvature singularity at any order on the $r=r_0$ horizon. Nevertheless we cannot extend the coordinate beyond the region $r\ge r_0$. The $r=r_0$ horizon with irrational $\Delta_+$ thus forms a natural boundary of the spacetime. Although there is no infinite tidal force on the horizon, this boundary is necessarily singular since the geodesics are incomplete on the horizon.

Physically, one may argue that an irrational number $\Delta_+$ can be approximated by a rational number $\Delta_+=p/q$, with $(p,q)$ being co-prime positive integers, in arbitrary accuracy. In this case, we can redefine the radial coordinate
\be
r=r_0 + \rho^q\,,\qquad\rightarrow\qquad (r-r_0)^{\fft{p}{q}}=\rho^p\,.
\ee
Under the new radial coordinate, the $r>r_0$ spacetime is described by $\rho> 0$, and the geodesics can be extended into the ``inside'' of the horizon where $\rho$ is negative.

  However, three distinct situations with equally good approximation can arise with the above
scenario. (1) $p$ is even and $q$ is odd; (2) $p$ is odd and $q$ is even; (3) $p$ and $q$ are both odd. In the first case, $(r-r_0)^{p/q}$ is an even function of $\rho$, {\it i.e.}, it is the same inside or outside the horizon, but the function $r$ is not. In the second case, the situation reverses, and $r$ remains the same inside or outside the horizon, but not for $(r-r_0)^{p/q}$. In the third case, both are odd functions of $\rho$. These three approximations of the irrational $\Delta_+$ therefore lead to three distinct insides of the horizon. Furthermore, the more accurate the fraction is to the irrational number, the larger the integers $q$ and $p$ must be, corresponding to larger number of branch cuts. An irrational power implies an infinite number of branch cuts; therefore, the global structures imply that the irrational power cannot be approximated by rational numbers in the black hole discussion.

\subsection{Origin of irrational $\Delta_+$}

In this paper, we considered the higher-derivative correction to rotating black holes, in order to restrict ourselves to a pure gravity discussion. We may therefore falsely accuse that the emerging of irrational $\Delta_+$ is a consequence of higher-derivative gravities. We now show that the vulnerability of the horizon regularity is actually an innate part of Einstein gravity. The general equation of motion can be expressed as
\be
G_{\mu\nu}(\Phi) + \alpha E_{\mu\nu}(\Phi)=0\,,
\ee
where $G_{\mu\nu}$ is the Einstein tensor, and $E_{\mu\nu}$ represents the differential operator associated with the higher-derivative correction. $\Phi$ denotes the metric functions to be solved. Perturbatively, we take $\Phi=\Phi_0 + \alpha \Phi_1$, with $G_{\mu\nu}(\Phi_0)=0$. At the linear $\alpha$ order, the perturbative equation is thus given by
\be
\delta G_{\mu\nu} (\Phi_1) = -E_1 (\Phi_0)\,.
\ee
Here $\delta G_{\mu\nu}$ represents the linearized Einstein tensor on the $\Phi_0$ background. Thus we see that in the perturbative analysis, the dynamics of the higher-derivative correction to the solution, {\it i.e.}~$\Phi_1$, is still governed by the Einstein's theory. The higher-derivative terms contribute only as a source to the linearized equation.

Therefore, the horizon regularity is not only vulnerable to higher-derivative corrections, but also to minimally-coupled matter fields in Einstein gravity. This may partially explain why exact solutions of rotating black holes in higher dimensions do not exist in Einstein-Maxwell theory in higher dimensions, since it is hard to imagine an analytic special function can give rises to the horizon structure \eqref{dfgend}. It is also worth commenting that since curvature tensors and their covariant derivatives are all non-singular at $r=r_0$, this vulnerability persists as long as we take perturbative approach, no matter how higher orders we consider in higher-derivative corrections.

\subsection{Fine-tuning away the horizon boundary}

Exact solutions of charged rotating black holes in $D=7$ do exist in gauged or ungauged supergravities \cite{Chong:2004dy,Chow:2007ts,Wu:2011gp,Chow:2011fh,Bobev:2023bxl} and there is no irrational horizon structure \eqref{dfgend} in these examples, even in the extremal limit.  Analogously, the non-analytic logarithmic terms in extremal five-dimensional rotating black hole found in \cite{Ma:2020xwi} do not arise in supergravity solutions constructed in \cite{Cvetic:2004hs,Chong:2005hr,Wu:2011gq} either. These examples contradict our earlier statements.

From our perturbative point of view, this can be explained that the source term $Q$ of \eqref{deltafgeneq} in these examples are fine-tuned in supergravities such that the divergent $c_{-2}$ vanishes at asymptotic infinity even when we set the coefficient $d_{\Delta_+}=0$. We expect that for the simpler Einstein-Maxwell theory in general odd dimensions, $d_{\Delta_+}$ coefficient will be necessarily turned on.

To validate the above arguments, we consider Einstein gravity with a cubic Riemann tensor extension, namely
\be
L_{\rm cubic}=\beta \Big(e_1 R^{\mu\nu}_{\ \ \rho\sigma}R^{\rho\sigma}_{\ \ \alpha\beta}R^{\alpha\beta}_{\ \ \mu\nu}
+e_2 R^{\mu\ \ \alpha}_{\ \ \nu\ \ \beta}R^{\nu\ \ \beta}_{\ \ \rho\ \ \gamma}R^{\rho\ \ \gamma}_{\ \ \mu\ \ \alpha}\Big).
\ee
We focus on $D=7$ dimensions, and we find that the source term in \eqref{deltafgeneq} is now given by
\bea
Q&=& \fft{96}{r^{10}}\bigg(30720 \left(12556 e_1+1301 e_2\right) \mu  \nu ^{10}+6531840 e_1 \mu ^2 r^{26}-311040 \left(263 e_1+13 e_2\right) \mu  \nu ^2 r^{24}\cr
&&-675 r^{22} \left(4 e_1 \left(34453 \mu ^3-57024 \nu ^4\right)-27 e_2 \left(19 \mu ^3+768 \nu ^4\right)\right)+1890 \left(442324 e_1+17957 e_2\right)\cr
&&\times \mu ^2 \nu ^2 r^{20}+30 \mu  r^{18} \left(28 e_1 \left(200305 \mu ^3-2409444 \nu ^4\right)-e_2 \left(36955 \mu ^3+5017716 \nu ^4\right)\right)\cr
&&-36 \nu ^2 r^{16} \left(e_1 \left(41738114 \mu ^3-39073536 \nu ^4\right)+3 e_2 \left(464483 \mu ^3-1297152 \nu ^4\right)\right)\cr
&&+\mu ^2 r^{14} \left(e_1 \left(4349474736 \nu ^4-39410300 \mu ^3\right)+e_2 \left(315875 \mu ^3+275913768 \nu ^4\right)\right)\cr
&&+6 \mu  \nu ^2 r^{12} \left(e_1 \left(55820180 \mu ^3-832631552 \nu ^4\right)+e_2 \left(1704055 \mu ^3-71862016 \nu ^4\right)\right)\cr
&&-240 r^{10} \big(e_1 \left(65500 \mu ^6+4001976 \mu ^3 \nu ^4-8136288 \nu ^8\right)+e_2 (-475 \mu ^6+243219 \mu ^3 \nu ^4\cr
&&-843048 \nu ^8)\big)+40 \mu ^2 \nu ^2 r^8 \left(e_1 \left(4254997 \mu ^3+28078336 \nu ^4\right)+32 e_2 \left(3601 \mu ^3+74962 \nu ^4\right)\right)\cr
&&-72 \mu  \nu ^4 r^6 \left(e_1 \left(9349998 \mu ^3+6333248 \nu ^4\right)+e_2 \left(490623 \mu ^3+669808 \nu ^4\right)\right)\cr
&&+20736 \left(60494 e_1+4387 e_2\right) \mu ^3 \nu ^6 r^4-128 \left(8742916 e_1+779963 e_2\right) \mu ^2 \nu ^8 r^2
\bigg).
\eea
In the extremal limit \eqref{extcond}, we have
\be
\delta f = \frac{1900 e_1-571 e_2}{5 r_0^4} \hat f_0(r) + d_{\Delta_+}\, (r-r_0)^{\Delta_+}\, \hat f_{\Delta_+}(r),
\ee
with the irrational $\Delta_+$ given in \eqref{d7delta}. Now we have an extra non-trivial free parameter $\xi=e_2/e_1$.  For generic $\xi$, the parameter $d_{\Delta_+}$ will be turned on appropriately for asymptotically-flat spacetime. In these cases, the horizon is the natural boundary of the spacetime. However, we can fine-tune the $\xi$ parameter so that we can turn off $d_{\Delta_+}$, so that the horizon is analytic. By numerical analysis, we find, up to six significant figures, that
\be
\xi\sim -2.51062.
\ee
In the linear perturbative solution, this quantity is independent of the size of the extremal black hole, namely the $r_0$ value.

\section{Conclusion}

In this paper, we obtained the leading-order perturbation of the extremal $D=7$ Ricci-flat rotating black holes with all equal angular momentum, in Einstein gravity extended with Riemann-squared term. The resulting mass-angular momentum relation is the same as one derived from the Reall-Santos procedure. We found that in the extremal limit, the perturbative solution of $1/g_{rr}$ took the form \eqref{dfhorizon} where $\Delta_+$ was an irrational number between half and one. We argued that the irrational $\Delta_+$ implied that the black hole inside of the original extremal black hole was destroyed by the perturbation and the horizon forms a natural boundary of spacetime.  All curvature invariants of the perturbation are regular on the horizon and hence there is no divergent tidal force; however, the horizon boundary is nevertheless singular since the geodesic is incomplete there. The origin of this singularity comes from the infinity number of branch cuts owing to the irrational power. We showed that this feature generally continued in higher odd dimensions, except in sporadically distributed special dimensions.

We demonstrated that this vulnerability of horizon regularity and black hole inside was not a consequence of higher-derivative corrections, but it was an innate part of Einstein gravity. In other words, horizon natural boundaries will generally arise in Einstein gravity with minimally-coupled matter in higher dimensions, unless the matter fields are fine-tuned as in supergravities. We also illustrated with a concrete example that we could also remove the non-analytic structure on the extremal horizon by fine-tuning the coupling constants in higher-derivative gravities.

The fact that the non-analytical terms that would generally exist in the near-horizon geometry of the rotating black holes in $D=5$ or $D=7$ actually do not arise in supergravities is tantalizing. Is this one of the criteria of a good quantum theory of gravity, which should be fine-tuned so as not to generate the horizon natural boundary? It is clearly problematic to understand the microscopic origin of the black hole entropy in the extremal limit when the inside of the horizon does not exist. It is thus worth investigating whether the supersymmetric higher-order correction would also protect the inside of the black holes.

We motivated ourself to study the horizon structure in the extremal limit of higher-dimensional rotating black hole by the fact that centrifugal force remains the same while gravity becomes weaker in higher dimensions. The extremal limit becomes a more strenuous balance between gravity and centrifugal force. We thus expected that the horizon became problematic under perturbation. However, it is curious to note that rational powers can nevertheless exist sporadically as indicated in Table 1, which provides a small number of counter examples to our expectations. Our analysis of the event horizon expansion of general non-extremal black holes in the appendix indicates that the metric functions are all analytic and infinitely differentiable. However, the situation with near-extremal cases where the non-extremality is of the order of $\alpha$ remains to be investigated further.

In this paper, for simplicity, we studied only the rotating black holes in odd dimensions with all equal angular momentum.  These solutions are all cohomogeneity-one metrics depending only on one radial variable. Although we expect that the arising of horizon natural boundaries will also occur in general rotating cases, the subjects require further investigation.

\section*{Acknowledgement}

We are grateful to Gary Gibbons, Yang Run-Qiu and Wang Zhao-Long for useful discussions. This work is supported in part by NSFC (National Natural Science Foundation of China) Grants No.~11935009 and No.~11875200.

\appendix

\section{Perturbed equation in general $D=2n+1$ dimensions}
\label{app:eom}

The four perturbed equations of $(\delta f, \delta h, \delta W, \delta \omega)$ are coupled second-order differential equations:
\bea
&&r^{2 D} \left(\left(D^2-2 D-3\right) \nu ^4 r^2+2 (D-5) \nu ^2 r^{D+1}+2 (D-3) r^{2 D}+(D-1)^2 \mu  \nu ^2 r^4\right)\delta W\cr
&&+2 \left(D^2-4 D+3\right) \left(r^D+\nu ^2 r\right)^2 \Big(\left(D^2-4 D-5\right) \nu ^4+\left(D^2-6 D+8\right) \mu ^2 r^4-2 (D^2\cr
&&-5 D+4) \mu  \nu ^2 r^2\Big) -(D-1)^2 \mu  \nu ^2 r^{2 D+4} \delta h +2 (D-3) r^D \left(r^D+\nu ^2 r\right)^2 \left((D-2) r^D-\nu ^2 r\right)\delta f\cr
&&-2 (D-1) \sqrt{\mu } \nu  r^{D+4} \left(r^D+\nu ^2 r\right)^2 \delta \omega'+r^{D+1} \left(r^D+\nu ^2 r\right) \left(2 (D-2) r^D+(D-3) \nu ^2 r\right)(r^D\cr
&&-\mu  r^3+\nu ^2 r) \delta h' -r^{2 D+1} \left(r^D+\nu ^2 r\right) \left(2 r^D+r^3 (\mu -D \mu )-(D-3) \nu ^2 r\right) \delta W'=0,\\
%%%%
&&-2 (D-3) (D-1) \left(r^D+\nu ^2 r\right) \Big(\left(D^2-6 D+8\right) \mu ^2 r^{D+4}-\left(D^2+4 D+3\right) \nu ^4 r^D\cr
&&+2 \left(D^2-1\right) \mu  \nu ^4 r^3-\left(D^2+4 D+3\right) \nu ^6 r-(D-4) D \mu ^2 \nu ^2 r^5\Big) +(D-1)^2 \mu  \nu ^2 r^{2 D+4}\delta h\cr
&&-r^{2 D}\left(\left(D^2-2 D-3\right) \nu ^4 r^2+2 (D-5) \nu ^2 r^{D+1}+2 (D-3) r^{2 D}+(D-1)^2 \mu  \nu ^2 r^4\right)\delta W\cr
&& -2 (D-3) r^D  \left(r^D+\nu ^2 r\right)^2 \left((D-2) r^D-\nu ^2 r\right) \delta f
+2 (D-1) \sqrt{\mu } \nu  r^{D+4} \left(r^D+\nu ^2 r\right)^2 \delta \omega '\cr
&&-r^{D+1} \left(r^D+\nu ^2 r\right) \left(2 (D-2) r^D+(D-3) \nu ^2 r\right)\left(r^D-\mu  r^3+\nu ^2 r\right) \delta f'\cr
&&-r^{2 D+1} \left(r^D+\nu ^2 r\right)  \Big(2 (D-1) r^D-(D+1) \mu  r^3 +3 (D-1) \nu ^2 r\Big) \delta W'\cr
&&-2 r^{2 D+2} \left(r^D+\nu ^2 r\right)\left(r^D-\mu  r^3+\nu ^2 r\right)\delta W''=0\,,\\
%%%%
%%%%
&&-4 (D-3) (D-1)^2 \left(r^D+\nu ^2 r\right) \left((D-4) \mu  r^2-(D+1) \nu ^2\right) \left(-\mu  r^{D+2}-\nu ^2 r^D+\mu  \nu ^2 r^3-\nu ^4 r\right)\cr
&&
-4 r^{2 D}\left(-\left(D^2-5 D+6\right) \nu ^4 r^2+4 (D-2) \nu ^2 r^{D+1}+(D-3) r^{2 D}-(D-1)^2 \mu  \nu ^2 r^4\right) \delta W\cr
&& -4 (D-1)^2 \mu  \nu ^2 r^{2 D+4}\delta h+4 (D-3) r^D \left(r^D+\nu ^2 r\right)^3 \delta f
-8 (D-1) \sqrt{\mu } \nu  r^{D+4} \left(r^D+\nu ^2 r\right)^2 \delta \omega'\cr
&&-2 r^{D+2} \left(r^D+\nu ^2 r\right)^2 \left(r^D-\mu  r^3+\nu ^2 r\right) \delta h''
+4 r^{2 D+2} \left(r^D+\nu ^2 r\right) \left(r^D-\mu  r^3+\nu ^2 r\right)\delta W''\cr
&&+4 r^{2 D+1} \left(r^D+\nu ^2 r\right) \left((D-2) r^D+(2 D-3) \nu ^2 r-\mu  r^3\right)\delta W' +r^{D+1} \left(r^D+\nu ^2 r\right) \cr
&&\times \left(-(D-1) \mu  r^{D+3}-(D-5) \nu ^2 r^{D+1}+2 r^{2 D}+(D-1) \mu  \nu ^2 r^4-(D-3) \nu ^4 r^2\right)\delta f'\cr
&&-r^{D+1} \left(r^D+\nu ^2 r\right)  \Big((D-3) \mu  r^{D+3}+(3 D-11) \nu ^2 r^{D+1}+2 (D-3) r^{2 D}\cr
&&-(D+1) \mu  \nu ^2 r^4+(D-5) \nu ^4 r^2\Big)\delta h'=0\,,\\
%%%%
&&-4 (D-3) (D-1)^2 \sqrt{\mu } \nu  r^{-D-2} \left((D-4) \mu  r^2-(D+1) \nu ^2\right)
-2 r^{-D} \left(r^D+\nu ^2 r\right)^2 \delta \omega ''\cr
&&-2 r^{-D-1} \left(r^D+\nu ^2 r\right) \left(D \left(r^D-\nu ^2 r\right)+2 \nu ^2 r\right) \delta \omega '+(D-1) \sqrt{\mu } \nu \delta f'-(D-1) \sqrt{\mu } \nu  \delta h'\cr
&&+2 D \sqrt{\mu } \nu \delta W'+2 \sqrt{\mu } \nu  r \delta W''=0\,.
\eea
By the standard procedure of eliminating variables, we can obtain a decoupled fourth-order differential equation for $\delta f$, given in section \eqref{sec:gend}. The functions $\delta h$ and $\delta \omega$ can be given as quadratures and $\delta W$ can be solved algebraically. They are
\bea
&&(D-2) \left(-8 (D-2) \mu  r^{D+3}+3 (D-1)^2 r^{2 D}-4 (D-2) \mu ^2 r^6+4 (D+1) \mu  \nu ^2 r^4\right) \delta h'=\cr
&&-2 (D-1) r^{-D-1} \Big(3 (D-3)^2 \left(D^3-7 D^2+14 D-8\right) \mu ^2 r^{D+4} -2 (D-3)^2(2 D^3-15 D^2\cr
&&+34 D-24) \mu ^3 r^7+4 (D+1)^2 \left(D^3-5 D^2-17 D+21\right) \nu ^6 r +2(-6 D^5+39 D^4+6 D^2+38 D\cr
&&-13) \mu  \nu ^4 r^3-2 \left(3 D^5-23 D^4+26 D^3+22 D^2-29 D+1\right) \mu  \nu ^2 r^{D+2} +(3 D^5-7 D^4-94 D^3\cr
&&-122 D^2+91 D+129) \nu ^4 r^D+4 \left(3 D^5-30 D^4+97 D^3-161 D^2+140 D-49\right) \mu ^2 \nu ^2 r^5\Big)\cr
&&-2 \left(D^3-6 D^2+11 D-6\right) r^{D-1} \left((D-5) r^D+2 \mu  r^3-4 \nu ^2 r\right) \delta f +(2-D) \Big(-2(2 D^2\cr
&&-15 D+23) \mu  r^{D+3}+2 \left(2 D^2-11 D+13\right) \nu ^2 r^{D+1}+\left(5 D^2-28 D+27\right) r^{2 D}\cr
&&+4 (7-2 D) \mu ^2 r^6+4 (D-6) \mu  \nu ^2 r^4+4 D \nu ^4 r^2\Big)\delta f'
+2 (D-2) r \left(-r^D+\mu  r^3-\nu ^2 r\right) \cr
&&\times \left((5 D-7) r^D+2 D \mu  r^3-2 D \nu ^2 r-14 \mu  r^3\right) \delta f'' -4 (D-2) r^2  \left(r^D-\mu  r^3+\nu ^2 r\right)^2 \delta f''', \label{eomh}\\
%%%%%
%%%%%
\cr
&&4(D-2) r^{2D} \delta W= 2 (D-1) \Big(-(D-7) (D-1) (D+1) \nu ^4-\left((D-4) (D-3) (D-2) \mu ^2 r^4\right)\cr
&&+2 (D-1) ((D-7) D+7) \mu  \nu ^2 r^2\Big) +(D-2) r^D \Big( (-2 (D-3) r^D +4 \nu ^2 r) \delta f\cr
&&-r \left(r^D-\mu  r^3+\nu ^2 r\right) \left(\delta f+\delta h\right) \Big), \label{eomw}\\
\cr
%%%%%%
%%%%%%
&&{2 (D-1) \sqrt{\mu } \nu  \left(r^D+\nu ^2 r\right)^2} \delta \omega'=  2 (D-3) (D-1) r^{-D-4} \left(r^D+\nu ^2 r\right)^2 \Big((D-5) (D+1) \nu ^4 \cr
&&+(D-4) (D-2) \mu ^2 r^4-2 (D-4) (D-1) \mu  \nu ^2 r^2\Big)-(D-1)^2 \mu  \nu ^2 r^D \delta h +\frac{2 (D-3)}{r^4}\cr
&&\times\left((D-2) r^D-2-\nu ^2 r\right) \left(r^D+\nu ^2 r\right)^2\delta f +r^{D-4}\Big(2 (D-5) \nu ^2 r^{D+1}+2 (D-3) r^{2 D}\cr
&&+(D-1)^2 \mu  \nu ^2 r^4+(D-3) (D+1) \nu ^4 r^2\Big)\delta W +\frac{\left(r^D+\nu ^2 r\right)}{r^3} \left(2 D r^D-4 r^D+D \nu ^2 r-3 \nu ^2 r\right)\cr
&&\times \left(r^D-\mu  r^3+\nu ^2 r\right) \delta h' -r^{D-3} \left(r^D+\nu ^2 r\right) \left(2 r^D-D \mu  r^3-D \nu ^2 r+\mu  r^3+3 \nu ^2 r\right) \delta W'.\label{eomo}
	\end{eqnarray}

\section{Near-Horizon geometry of non-extremal black holes}
\label{app:nhne}

For non-extremal black holes, we can solve for $\mu$ in terms of $\nu$ and $r_0$, i.e.
\be
\mu=r_0^{D-3}+\frac{\nu ^2}{r_0^2}\,.
\ee
To obtain the behavior of  $\delta f$ in the near-horizon region, we can assume an ansatz $\delta f = (r-r_0)^\gamma \hat f$. We find that at the leading order of $(r-r_0)$, we have
\be
(\gamma-1)\gamma (\gamma+1)^2 \hat f =Q_0\, (r-r_0)^{2-\gamma}\,,
\ee
where
\bea
Q_0&=&\frac{2 (D-1) r_0^{-D-4}}{3 (D-2) \left((D-3) r_0^D-2 \nu ^2 r_0\right)^2}\Big(
2 \left(119 D^3-1406 D^2+1739 D-336\right) \nu ^6 r_0^3\cr
&&-\left(289 D^4-3528 D^3+11220 D^2-11684 D+3519\right) \nu ^4 r_0^{D+2}\cr
&&-(D-3)^2 \left(9 D^4-86 D^3+289 D^2-406 D+200\right) r_0^{3 D}\cr
&&+2 \left(52 D^5-685 D^4+3162 D^3-6617 D^2+6488 D-2436\right) \nu ^2 r_0^{2 D+1}
\Big).
\eea
The modes associated with the double roots $\gamma=-1$ is clearly unacceptable for the regular horizon, we thus have
\be
\delta f = d_0\, \hat f_0(r) + d_1\, \hat f_1(r) + \fft{Q_0}{18}\, \hat f_s(r)\,.
\ee
Here $d_0$ and $d_1$ are constants to be determined so that the $c_2$ term at asymptotic infinity will not be generated. The last term above represents the contribution from the source. All $f_i(r)$ functions are analytic with Taylor expansions of the form $\hat f\sim 1 + \#_1 (r-r_0) + \#_2 (r-r_0)^2 + \cdots$.

At the first sight, there appears to have two many parameters on the horizon, namely $(r_0,\nu, d_0, d_1)$. However, in a typical discussion of perturbative solutions, we need to hold the mass and angular momentum fixed. In our case, we have already hold the $r_0$ fixed, we can now hold either mass or angular momentum fixed. In this case, we need fix one combination of the coefficients $(\nu, d_0, d_1)$. Furthermore, we need fine-tune these parameters so that the asymptotic divergent $c_{-2}$ term vanishes, and this fix another combination, leaving only one combination, together with $r_0$.

We present the near-horizon structure at low-lying orders of the Taylor expansion for all the functions only in $D=7$,
\bea
&\delta f&=d_0+\left(r-r_0\right) d_1+\left(r-r_0\right)^2 \Bigg(-\frac{16 \left(1815 r_0^{18}-2037 r_0^{12} \nu ^2-1397 r_0^6 \nu ^4+1015 \nu ^6\right)}{45 r_0^{10} \left(2r_0^6-\nu ^2\right)^2}\cr
&&+\frac{4 r_0^4 \left(31 r_0^6+4 \nu ^2\right) d_0}{9 \left(2 r_0^6-\nu ^2\right)^2}-\frac{\left(74 r_0^6+47 \nu ^2\right) d_1}{54r_0( 2r_0^6- \nu ^2)}\Bigg)+\left(r-r_0\right)^3 \Bigg(\cr
&&+\frac{2 \left(95475 r_0^{24}-150483 r_0^{18} \nu ^2-64897 r_0^{12} \nu ^4+91711 r_0^6 \nu ^6-15910 \nu ^8\right)}{45 r_0^{11}\left(2 r_0^6-\nu ^2\right)^3}\cr
&&-\frac{r_0^3 \left(605 r_0^{12}+637 r_0^6 \nu ^2+32 \nu ^4\right) d_0}{18 \left(2 r_0^6-\nu ^2\right)^3} +\frac{\left(547 r_0^{12}+587r_0^6 \nu ^2+40 \nu ^4\right) d_1}{108 r_0^2\left(2 r_0^6-\nu ^2\right)^2}\Bigg),\nn\\
%%%%%%			
&\delta h& =h_0+\left(r-r_0\right) \left(-\frac{d_1}{3}-\frac{8 d_0 r_0^5}{2 r_0^6-\nu ^2}+\frac{32 \left(7 \nu ^4-18 \nu ^2 r_0^6+15 r_0^{12}\right)}{5 r_0^9 \left(2 r_0^6-\nu ^2\right)}\right)\cr
&&+\left(r-r_0\right)^2 \Bigg(-\frac{16 \left(-788 \nu ^6+3037 \nu ^4 r_0^6-3540 \nu ^2 r_0^{12}+1635 r_0^{18}\right)}{45 r_0^{10} \left(2 r_0^6-\nu ^2\right)^2}+\frac{76 d_0 r_0^4 \left(\nu ^2+r_0^6\right)}{9 \left(2 r_0^6-\nu ^2\right)^2}\cr
&&+\frac{d_1 \left(13 \nu ^2-122r_0^6\right)}{54 r_0\left(2 r_0^6-\nu ^2\right)}\Bigg)+\cdots\,,\\
%%%%%
&\delta W&=-\frac{36 \left(-2 \nu ^4+3 \nu ^2 r_0^6+5 r_0^{12}\right)}{5 r_0^{14}}+d_0 \left(-2+\frac{\nu ^2}{r_0^6}\right)+\left(r-r_0\right) \Bigg(\frac{4}{3} d_1			\left(-2+\frac{\nu ^2}{r_0^6}\right)\cr
&&+\frac{d_0 \left(-6 \nu ^2+4 r_0^6\right)}{r_0^7}+\frac{8 \left(-167 \nu ^4+108 \nu ^2 r_0^6+195 r_0^{12}\right)}{5r_0^{15}}\Bigg)+\left(r-r_0\right)^2 \Bigg(\frac{6 d_1 \left(-\nu ^2+r_0^6\right)}{r_0^7}\cr
&&-\frac{3 d_0 \left(7 \nu ^4-16 \nu ^2 r_0^6+16 r_0^{12}\right)}{r_0^8	
\left(2 r_0^6-\nu ^2\right)}-\frac{12 \left(671 \nu ^6-1810 \nu ^4 r_0^6+489 \nu ^2 r_0^{12}+810 r_0^{18}\right)}{5 r_0^{16} \left(2 r_0^6-\nu ^2\right)}\Bigg)+\cdots\,,\cr
%%%%%			
&\delta \omega&=\omega_{0}+\left(r-r_0\right) \Bigg(-\frac{3 \nu  h_{t_0} r_0^4}{\left(\nu ^2+r_0^6\right)^{3/2}}-\frac{2 d_1 \left(-5 \nu ^4+8 \nu ^2 r_0^6+4 r_0^{12}\right)}{9 \nu  r_0\left(\nu ^2+r_0^6\right)^{3/2}}\cr
&&+\frac{d_0 r_0^4 \left(-29 \nu ^4+8 \nu ^2 r_0^6+10 r_0^{12}\right)}{3 \nu  \left(\nu ^2+r_0^6\right)^{5/2}} +\frac{4 \left(-604 \nu^6-544 \nu ^4 r_0^6+912 \nu ^2 r_0^{12}+525 r_0^{18}\right)}{15 \nu  r_0^{10} \left(\nu ^2+r_0^6\right)^{3/2}}\Bigg)+\cdots\,.\nn
\eea
Note that all the expansion powers are integers and the functions therefore are all analytic and infinitely differentiable. Taking an extremal limit, corresponding to setting $\nu=\sqrt{2} r_0^3$, is singular in this expansion. In the above expansion, we have assumed that $(\nu^2-2r_0^6)/r_0^4\gg \alpha$. The near-extremal case with $(\nu^2-2r_0^6)/r_0^4\sim \alpha$ remains further study.

\section{Near-horizon expansion of the $D=7$ extremal black hole}
\label{app:nhe}

In section \ref{sec:sub-nhne}, we give the near-horizon expansion of the perturbative function $\delta f$ for the $D=7$ extremal rotating black hole. Here, we give the these expansions for the remaining functions:
\bea
\delta h &=& \Big(\hat h_0 -\frac{15704 \left(r-r_0\right)}{15 r_0^3}
+\frac{68164 \left(r-r_0\right)^2}{315 r_0^4}+ \cdots\Big) - d_{\Delta_+}\, \left(r-r_0\right)^{\frac{1}{2} \left(\sqrt{21}-3\right)}
\Big(\ft{1}{6} \left(\sqrt{21}+3\right)\cr
&&+\frac{\left(11 \sqrt{21}+24\right) \left(r-r_0\right)}{90 r_0}+\frac{\left(27551 \sqrt{21}-121011\right) \left(r-r_0\right)^2}{36720 r_0^2} + \cdots \Big),\nn\\
\delta W &=& \Big(-\frac{108}{5 r_0^2} -\frac{456 \left(r-r_0\right)}{r_0^3} -\frac{27444 \left(r-r_0\right){}^2}{5 r_0^4} + \cdots\Big) + d_{\Delta_+}\,\left(r-r_0\right)^{\frac{1}{2} \left(\sqrt{21}-1\right)}\Big(\cr
&&-\frac{3 \left(\sqrt{21}+3\right)}{2 r_0} +\frac{3 \left(31 \sqrt{21}+99\right) \left(r-r_0\right)}{20 r_0^2} -\frac{\left(57947 \sqrt{21}+114813\right) \left(r-r_0\right){}^2}{4080 r_0^3}
+ \cdots\Big),\nn\\
%%%%%
\delta \omega &=& \Big(\hat \omega _0-\frac{\left(\sqrt{\frac{2}{3}} \left(5 \hat h_0 r_0^2+1082\right)\right)\left(r-r_0\right)}{5 r_0^4}-\frac{\left(5 \hat h_0 r_0^2+11226\right)\left(r-r_0\right)^2}{5 \left(\sqrt{6} r_0^5\right)}+\cdots\Big)\cr
&&+d_{\Delta_+}\left(r-r_0\right)^{\frac{1}{2} \left(\sqrt{21}-1\right)}\Big(
-\frac{2 \sqrt{5+\sqrt{21}}}{3 r_0^2}+\frac{\left(33 \sqrt{14}-41 \sqrt{6}\right) \left(r-r_0\right)}{90 r_0^3}\cr
&&+\frac{\left(47953 \sqrt{6}+261 \sqrt{14}\right) \left(r-r_0\right)^2}{18360 r_0^4}+\cdots\Big).
\eea
We have two new integration constants $(\hat h_0, \hat \omega_0)$. The former should be chosen such that the speed of light at the asymptotic infinity remains unit, whilst the latter should be chosen so that the asymptotic spacetime is non-rotating.

\end{document}